\documentclass[9pt,twocolumn,twoside]{opticajnl}
\journal{opticajournal} 
\usepackage[utf8]{inputenc}
\DeclareUnicodeCharacter{03BB}{$\lambda$}
\DeclareUnicodeCharacter{03C0}{$\pi$}
\setboolean{shortarticle}{true}

\usepackage{subcaption}
\makeatletter
\renewcommand*\subcaption@label{\caption@withoptargs\subcaption@@label}
\makeatother
\usepackage{stfloats}
\usepackage{multirow}
\usepackage{siunitx}
\usepackage{physics}
\usepackage{pifont}
\DeclareSIUnit\baud{Baud}

\usepackage{lineno}

\title{ Coexistence of continuous-variable quantum key distribution and classical data over 120-km fiber}

\author[1, *]{Adnan A.E. Hajomer}
\author[2]{Ivan Derkach}
\author[2]{Vladyslav C. Usenko}
\author[1]{Ulrik L. Andersen}
\author[1,*]{Tobias Gehring}

\affil[1]{Center for Macroscopic Quantum States (bigQ), Department of Physics, Technical University of Denmark, 2800 Kongens Lyngby, Denmark}
\affil[2]{Department of Optics, Faculty of Science, Palacky University, 17. listopadu 12, 771 46 Olomouc, Czech Republic}
\affil[*]{Corresponding authors: aaeha@dtu.dk, tobias.gehring@fysik.dtu.dk}

\begin{abstract}
Integrating quantum key distribution (QKD) with classical data transmission over the same fiber is crucial for scalable quantum-secured communication. However, noise from classical channels limits QKD distance. We demonstrate the longest-distance continuous-variable QKD (CV-QKD) over 120 km (20 dB loss) in the asymptotic regime, and over 100 km (17 dB loss) in the finite-size regime, both coexisting with a fully populated coarse wavelength division multiplexing system. Natural mode filtering of the local oscillator and phase noise mitigation enabled this without additional filtering or wavelength reallocation. Benchmarking against a commercial discrete-variable QKD system and considering finite-size effects confirms the feasibility of CV-QKD  as a plug-and-play solution for typical 80–100 km long-haul optical networks. Our results set a record fiber distance for CV-QKD, showing its potential for cost-effective, large-scale deployment in existing network infrastructure.
\end{abstract}

\setboolean{displaycopyright}{false} 

\begin{document}

\maketitle

Quantum key distribution (QKD) has emerged as a leading quantum technology, reaching a maturity level suitable for real-world deployment and commercialization \cite{pirandola2020advances}. It is broadly categorized into discrete-variable (DV) and continuous-variable (CV) QKD~\cite{usenko2025continuous}, both of which have demonstrated secure key generation over long distances in dedicated optical fibers \cite{boaron2018secure, zhang2020long, liu2023experimental, hajomer2024long}. However, large-scale deployment requires QKD to coexist with classical data transmission within the same optical fiber channel to minimize infrastructure costs associated with deploying and maintaining dark fibers.

A promising approach to achieving this coexistence is wavelength division multiplexing (WDM), where QKD and classical signals use different frequency bands but otherwise share the same fiber infrastructure. However, the fundamental challenge lies in the extreme fragility of quantum states, which are highly susceptible to noise photons from co-propagating classical signals despite being in different frequency bands. This noise primarily stems from the significant power imbalance between classical and quantum signals, as well as Raman scattering and other nonlinear fiber effects, which lowers the attainable secure key rate as well as the maximum transmission distance of QKD systems.

Extensive efforts have been made to mitigate classical noise and enable the coexistence of QKD and classical signals. In DV QKD, noise reduction strategies typically involve temporal filtering using gated detectors, spectral filtering, and operating within specific wavelength windows (e.g., around 1310 nm) to minimize Raman-induced noise~\cite{townsend1997simultaneous, eraerds2010quantum, patel2012coexistence, wang2017long, gavignet2023co}. While effective, these constraints can limit seamless QKD integration into existing telecom networks, requiring dedicated wavelength allocation or the implementation of strong filtering mechanisms.

In contrast, CV-QKD offers a promising alternative by encoding and decoding quantum information using standard telecom components, such as quadrature modulators and coherent detectors, with a local oscillator (LO) facilitating the detection process. A key advantage of CV-QKD is that the LO inherently acts as a mode filter, improving resilience to classical noise and therefore enabling seamless integration into existing fiber networks. Previous demonstrations of CV-QKD coexistence with classical channels have primarily focused on analyzing different sources of noise and increasing the number of multiplexed classical signals~\cite{kumar2015coexistence,eriksson2019wavelength,eriksson2019crosstalk, karinou2018toward, milovanvcev2021high}. In particular, these works have been limited by short secure transmission distances. 

In this work, we report the longest-distance demonstration of  CV-QKD coexisting with a fully populated coarse wavelength division multiplexing (CWDM) system. Specifically, we demonstrate co-propagation 
of CV-QKD and classical data transmission channels, each with about 1 mW launch power provided by off-the-shelf 10G small form-factor pluggable (SFP) transceivers, over 100 km and 120 km of ultra-low-loss optical fiber, corresponding to total channel losses of 17 dB and 20 dB, respectively. Secure keys are generated in the finite-size regime at 100 km, while asymptotic keys are achieved at 120 km. This record transmission distance is enabled by the intrinsic mode-filtering properties of the LO and the optimization of modulation variance to suppress phase noise-induced excess noise. We further validate the effectiveness of mode filtering by comparing excess noise levels with and without classical channels. Finally, we benchmark our CV-QKD system against a commercial DV QKD system operating over the same link, demonstrating that CV-QKD provides a viable, plug-and-play solution for integrating QKD into typical long-haul optical networks spanning 80–100 km.

\begin{figure*}
    \centering
    \includegraphics[width=0.75\linewidth]{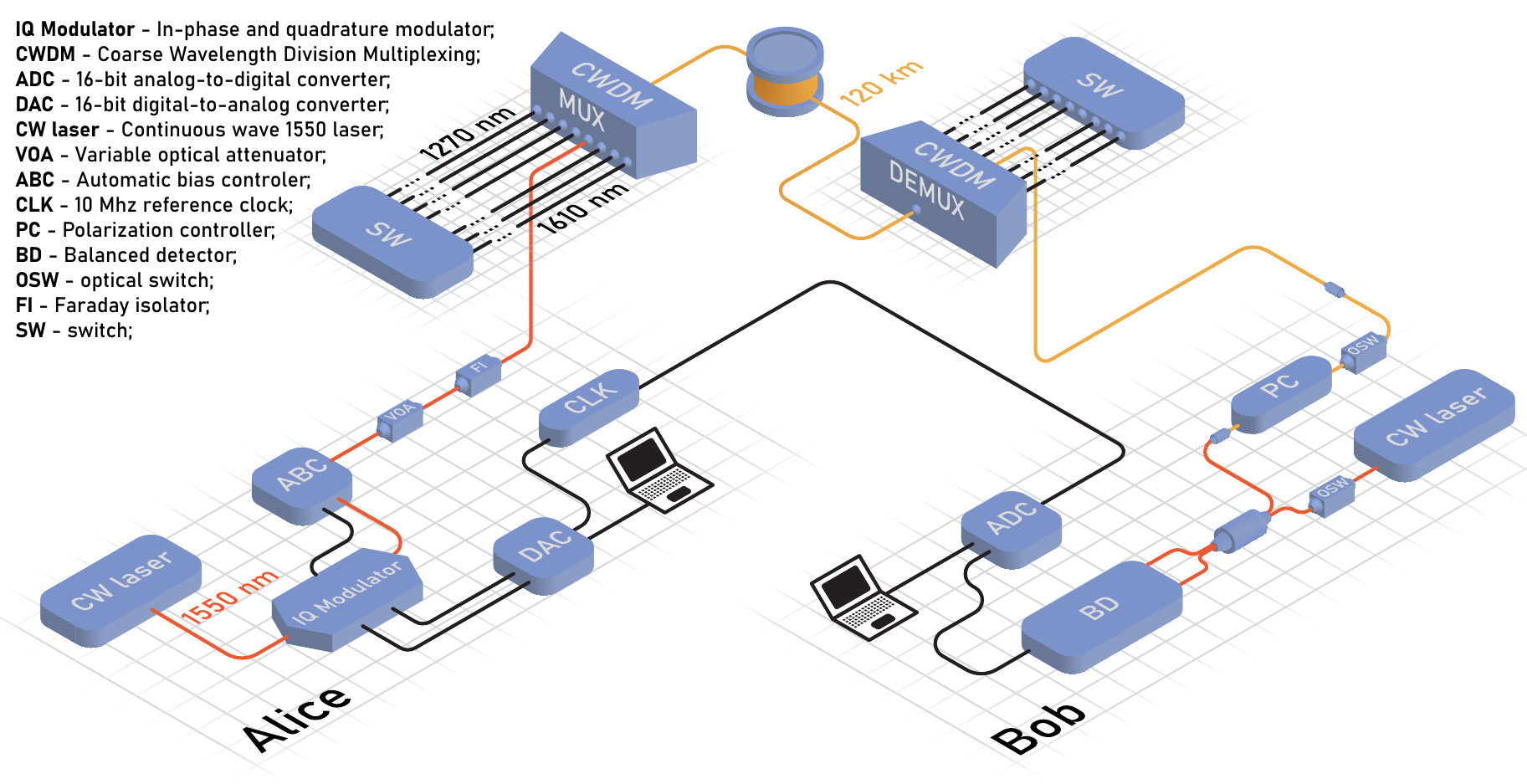}
    \caption{ \textbf{ Experimental setup for CV-QKD coexisting with a fully populated CWDM system}.}
    \label{fig:scheme}
\end{figure*}

Figure~\ref{fig:scheme} shows our long-distance local local-oscillator (LLO) CV-QKD system~\cite{qi2015generating} based on Gaussian-modulated coherent states consisting of a transmitter (Alice), a receiver (Bob), and a CWDM system.

At the transmitter, a continuous-wave (CW) laser at 1550 nm, with a narrow linewidth of approximately 100 Hz, was employed as the optical source. Coherent states were prepared using an in-phase and quadrature (IQ) modulator, driven by a 16-bit digital-to-analog converter (DAC) operating at 1 Gsample/s. The optical single-sideband modulation was realized by precisely adjusting the bias voltages using an automatic bias controller (ABC). A variable optical attenuator (VOA) fine-tuned the modulation variance of the transmitted Gaussian distribution of coherent states. To prevent back-reflections and mitigate Trojan-horse attacks, a Faraday isolator was introduced at the transmitter’s output. 

Quantum symbols were derived from Gaussian-distributed random numbers generated by a quantum random number generator (QRNG) based on vacuum fluctuations~\cite{Gehring2021qrng}. These symbols ($\alpha_i = x_i + ip_i$) were mapped onto amplitude and phase quadrature in phase space. The system's symbol rate was 125 MBaud. The quantum symbols were up-sampled to the DAC sampling rate of 1 Gsample/s and pulse shaped using a root-raised cosine (RRC) filter with a roll-off factor of 0.2. The quantum signal was frequency-shifted to 120 MHz for single-sideband modulation and multiplexed with a pilot tone at 220 MHz for carrier phase recovery.

After the CV-QKD transmitter, the quantum signal at 1550 nm was multiplexed with CWDM channels using a multiplexer (Mux). Each CWDM channel utilized a commercial 10 Gbit/s SFP+ transceiver from Nexgen with a launch power of about 1~mW (0 dBm), rated for 21 to 24 dB channel attenuation. To construct full-duplex system, the SFP transceivers were installed in two ARISTA DCS-7050S-64 data center Ethernet switches, each equipped with 48 1/10 GbE SFP ports. The CWDM system, spanning wavelengths from 1270 nm to 1610 nm with 20 nm channel spacing, was fully populated with seventeen classical channels. Following the CWDM Mux, the quantum and classical signals were transmitted through a 120 km ultra-low-loss fiber (TeraWave® SCUBA 125 Ocean Optical Fiber) with an attenuation of 0.146 dB/km at 1550 nm. The total fiber channel loss was measured at approximately 18 dB, primarily due to a mode field diameter mismatch between the SMF-28 fiber pigtail and SCUBA 125 fiber. At the receiver, the classical and quantum signals were separated using a demultiplexer (DeMux). The throughput of the classical channels was evaluated using the open-source tool iPerf, achieving an average data rate of 9.81 Gbit/s per channel. The Mux and DeMux introduced at the 1550 nm port an insertion loss of approximately 1.5 dB, which was considered part of the total channel loss.

At the receiver, quantum state measurements were performed via radio-frequency (RF) heterodyne detection. An independent CW laser--of the same type as the transmitter laser, with a 100 Hz linewidth--operated as the LLO. This laser was free-running relative to Alice’s laser and maintained a frequency offset of approximately 240 MHz. The quantum signal’s polarization was aligned to match that of the LO using a manual polarization controller (PC). Interference between the quantum signal and the LO was detected via a balanced detector (BD)  with a bandwidth of ~250 MHz. The detector's output was digitized by a 16-bit analog-to-digital converter (ADC) operating at 1 Gsample/s. A 10 MHz reference clock was used to ensure precise synchronization between the DAC and ADC. Two optical switches (OSW) were added to the quantum signal and LO path to perform vacuum noise and electronic noise calibration. 

The received signal was subjected to extensive digital signal processing (DSP) to recover quantum symbols. To speed up the DSP, the measurements were divided into frames, each with $10^7$ DAC samples. Initially, a frequency-domain equalizer (whitening filter) was applied to remove auto-correlation and maintain the statistical independence of quantum symbols. To estimate the frequency offset between the two lasers, the pilot tone was extracted using a 1 MHz bandpass filter. A linear fit was used to estimate the frequency offset after extracting the phase profile using the Hilbert transform. An unscented Kalman filter was used for phase estimation~\cite{chin2021machine}. The quantum signal was then shifted to the baseband and corrected for the phase noise. Finally, after matched RRC filtering and down-sampling, the quantum symbols were reconstructed.

Following quantum state reconstruction, classical post-processing was applied to extract secure keys. This included information reconciliation, parameter estimation, and privacy amplification. Information reconciliation was carried out using multidimensional reconciliation based on multi-edge-type (MET) low-density parity-check (LDPC) codes with a fixed rate of 0.01 and a codeword length of 819200 bits~\cite{mani2021multiedge}, achieving a reconciliation efficiency ($\beta$) of 93.73\% and a frame error rate (FER) of 30\%.
\begin{table}[t]
\centering
\caption{\textbf{Experimental parameters for different fiber lengths}. D: distances,  $V_M$: modulation variance,  
$\eta$: Untrusted efficiency, $\xi$: Excess noise (at channel output after back-propagating with the trusted efficiency), $\tau$: trusted efficiency, $V_{el}$: Trusted detection noise, $N$: block size.}
\resizebox{1\hsize}{!}{
\begin{tabular}{cccccccc}
\hline
\bf D, km & \bf $V_M$, SNU & \bf $\eta$ & $\xi$, mSNU & \bf $\tau$ & \bf $V_{el}$, mSNU & $N$ & \bf CWDM Channels\\
\hline
100 & 9.41 & 0.0180 & 0.714 & 0.685 & 19.62 & $1\times10^8$ & OFF \\
100 & 9.41 & 0.0184 & 0.760 & 0.685 & 19.25 & $1\times10^8$ & ON \\
100 & 9.41 & 0.0183 & 0.712 & 0.685 & 19.26 & $1.6\times10^9$ & ON \\
120 & 4.71 & 0.0096 & 0.444 & 0.685 & 18.99 & $1\times10^8$ & ON \\
\hline
\end{tabular}
}
\label{tab:1}
\end{table}

We evaluate the performance of the integrated CV-QKD system based on the no-switching protocol~\cite{weedbrook2004quantum} with CWDM channels using two key metrics: the secret key rate (SKR) and excess noise. Experiments were conducted under two conditions: (i) with classical channels turned OFF, serving as a reference, and (ii) with classical channels turned ON. Table~\ref{tab:1} summarizes the experimental parameters for fiber channels of 100 km and 120 km. In both cases, Alice generates coherent states with a modulation variance $V_M$ of 9.41 SNU for the 100 km link and 4.71 SNU for the 120 km link. Due to additional insertion losses introduced by the CWDM system and optical switches, the effective loss in the 100 km and 120 km channels exceeds the physical loss of fiber channels.  

\begin{figure}[h]
    \centering
    \includegraphics[width=.9\linewidth]{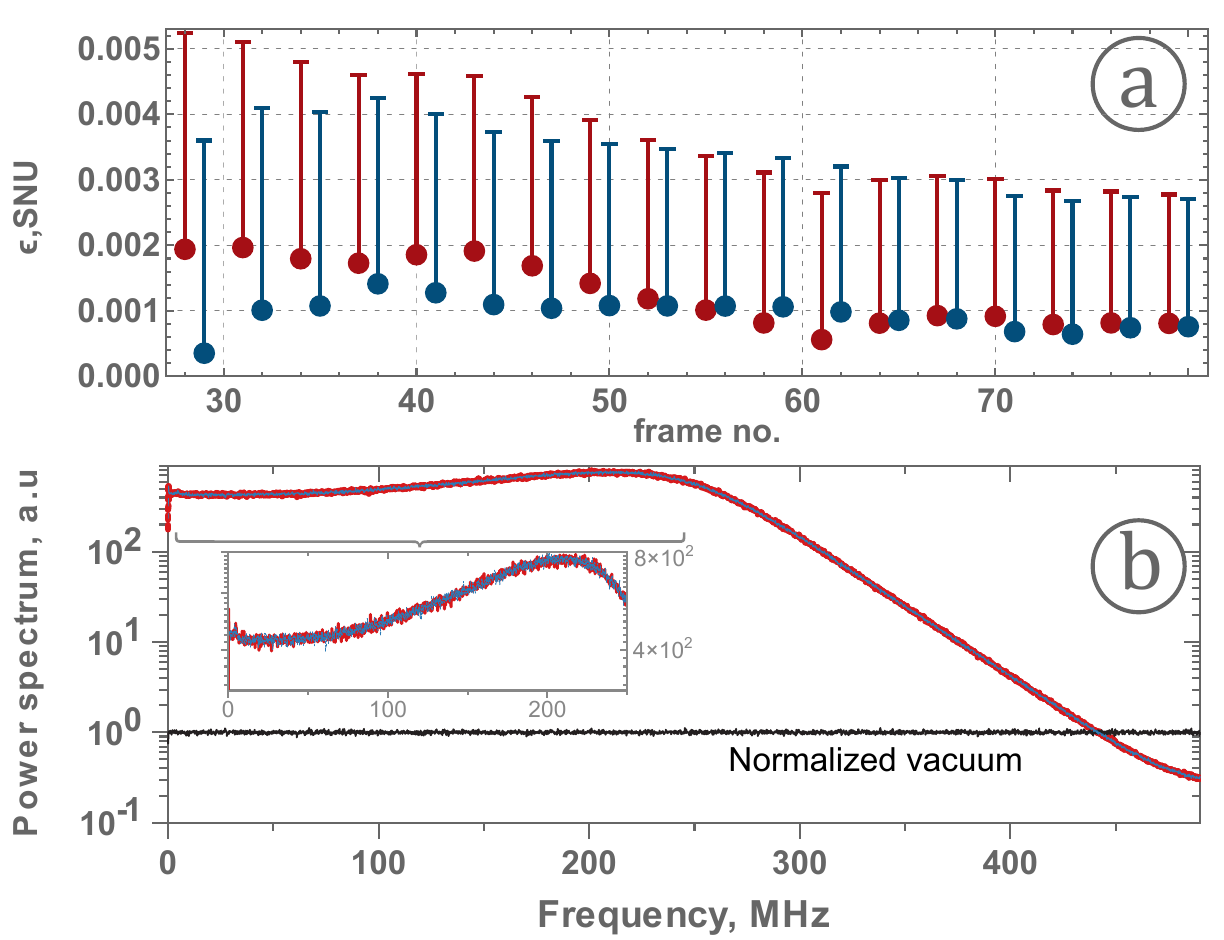}
    \caption{(\textbf{a}) Cumulative excess noise (at the channel output after back-propagating with $\tau$) after each measurement frame (each of $\approx 1.2\times10^6$ symbols) when classical channels are turned \textit{OFF} (red) or \textit{ON} (blue). Error bars are evaluated based on Gaussian confidence intervals for given amount of acquired data and protocol failure probability of $\epsilon=10^{-10}$ \cite{ruppert2014long} \textcolor{black}{and limited detector efficiency $\tau$}. (\textbf{b}) Vacuum noise power with (blue) and without (red) classical channels is identical across broad range of frequencies. }
    \label{fig:counts}
\end{figure}

To quantify the impact of classical channels on excess noise, we compared measurements taken with and without their presence. Figure~\ref{fig:counts}(a) shows the excess noise at the quantum channel output after back-propagating with $\tau$ as a function of the accumulated number of symbols per frame ($\approx 1.2\times10^6$ symbols) over the 100 km fiber channel. The average measured excess noise is 0.714 mSNU with classical channels off and 0.760 mSNU with them on. Despite the slight increase, the overlapping error bars indicate that this variation is statistically insignificant, demonstrating that the presence of classical channels has a negligible effect on the excess noise performance of CV-QKD. Compared to Ref.~\cite{hajomer2024long}, the introduction of a CWDM system adds approximately 2~dB of insertion loss and necessitates operating at a higher modulation variance, which leads to slightly increased excess noise but enables improved reconciliation efficiency and lower FER.  

Further validation comes from analyzing the power spectrum of vacuum noise, comparing the case with classical channels ON to that of true vacuum noise (i.e., classical channels OFF). This comparison is illustrated by the red and blue traces in Fig.~\ref{fig:counts}(b), which completely overlap. This is further confirmed by the normalized vacuum noise (black trace) plotted on a logarithmic scale, remaining at a value of 1 across the detector bandwidth, indicating no observable difference between the two scenarios. These results demonstrate that the classical channels do not introduce significant excess noise, primarily due to the natural mode filtering effect of the LO.  

\begin{figure*}[t]
    \centering
    \includegraphics[width=0.9\linewidth]{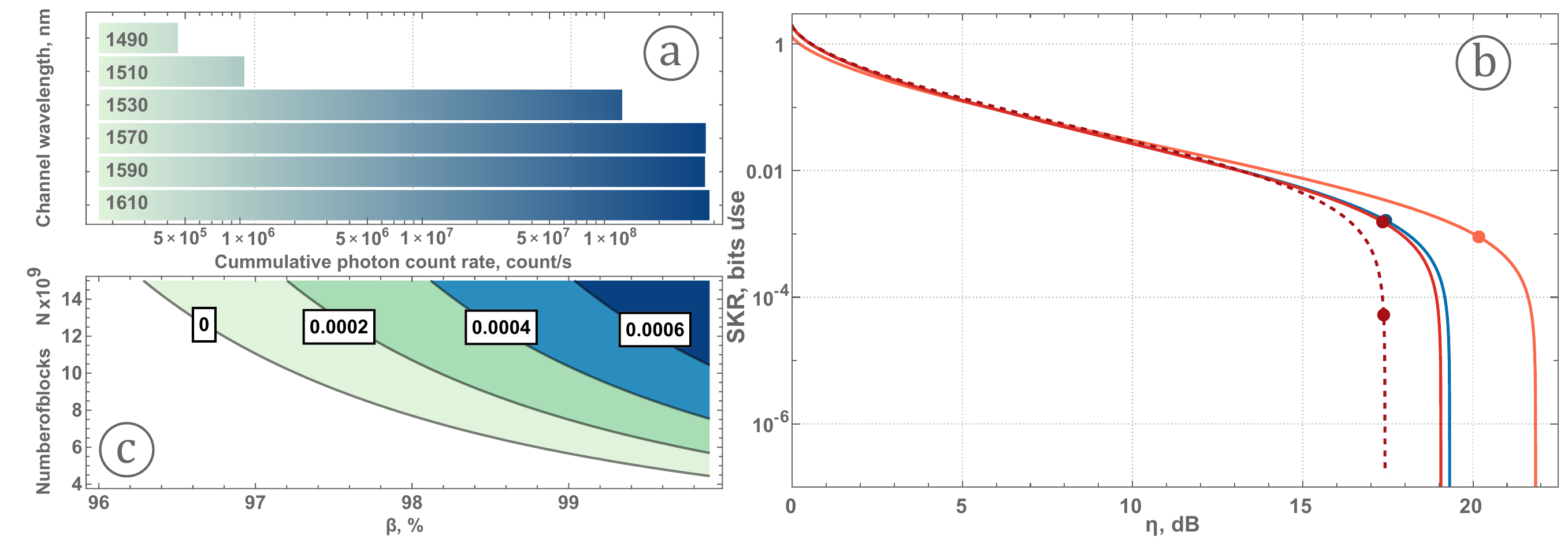}
    \caption{(a) Cumulative photon count rate (in count/s) at 1550 nm port. (b) Secure key rate (in bits/channel use) dependency on the channel loss (in dB). Theoretical secure transmittance is shown for asymptotic (solid) and finite-size (dashed) regime, without classical channels (blue lines) and with classical channels (red). Experimental measurement results are indicated by red circles for measurements with classical channels and a blue circle for measurements without classical channels. (c) \textcolor{black}{Secure} key rate (in bits/channel use) as a function of block size and information reconciliation efficiency.}
    \label{fig:100 km}
\end{figure*}

To verify the LO's mode-filtering properties, we assess leakage from classical channels by measuring photon noise at the 1550 nm port using superconducting nanowire single-photon detectors (SNSPDs), with the MUX and DEMUX connected via a short fiber patch cord. Figure~\ref{fig:100 km}(a) presents the cumulative photon count rate after enabling classical channels within the 1550 nm window and subtracting the dark count of 126 count/s. The results indicate that the dominant noise contribution originates from adjacent channels at 1530 nm and 1570 nm, as the cumulative photon count rate  saturates upon activating these two channels. Despite this observed photon leakage, the excess noise of our CV-QKD system remains unaffected (Fig.~\ref{fig:counts} (a)), further confirming the LO’s mode-filtering capability and the system’s robustness against interference from classical channels. 

Additionally, we have seen no effect of spontaneous Raman scattering noise, as previous studies~\cite{kumar2015coexistence,karinou2018toward} have shown that it decreases with both distance and launch optical power. This conclusion is further supported by our excess noise measurements, which confirm that Raman scattering is negligible at the observed total loss.

The SKR (in bits per channel use) quantifies the information advantage of Alice and Bob over Eve~\cite{pirandola2020advances}, as follows:
\begin{equation}
    R=\max\left[0,\beta \left[I(A:B)_x+I(A:B)_p \right]^{\epsilon_{PE}} -\chi^{\epsilon_{PE}}_E-\Delta(n)^{\epsilon_{PA},\bar{\epsilon}}\right],
\end{equation} 
where $I(A:B)_{x(p)}^{\epsilon_{PE}}$ is the mutual information between Alice and Bob in respective quadrature, scaled down by the reconciliation efficiency $\beta \in \left[0,1\right]$. \textcolor{black}{Here $\epsilon$ indices specify respective failure probabilities associated with each term. }The upper bound on the eavesdroppers information is given by $\chi_E=S(ABCD)-S(ACD|B)$, with $S(ABCD)$ being the von Neumann entropy of the shared trusted state (where modes $C$ and $D$ contain the purification of the electronic receiver noise), and $S(ACD|B)$ is the entropy of the state conditioned by the heterodyne measurement conducted at the receiver. The correction term $\Delta(n)$ is related to the security of privacy amplification \cite{leverrier2010finite} and depends on the number of symbols $n$ used for key generation, privacy amplification failure probability $\epsilon_{PA}$ and smoothing parameter $\bar{\epsilon}$. 
Both $I(A:B)_{x(p)}^{\epsilon_{PE}}$ and $\chi_E^{\epsilon_{PE}}$ depend on the worst-case estimators of channel parameters \textcolor{black}{(and limited detector efficiency $\tau$)}, that are estimated correctly with probability $1 - \epsilon_{PE}$. In the asymptotic regime $\Delta(n)$ is absent, and channel parameters are presumed to be true, disregarding the estimation confidence region determined by $\epsilon_{PE}$ and absolute values of respective parameters. For further details see Refs.\cite{leverrier2010finite} and \cite{zenodo-code}. 

Figure~\ref{fig:100 km}(b) illustrates the SKR obtained from both simulations (solid and dashed lines) and experimental results (data points), considering both asymptotic (solid lines) and finite-size (dashed lines) regimes as a function of channel loss. To ensure a fair comparison between the classical channels ON and OFF cases, we assume the same reconciliation efficiency $\beta$ of 96\%~\cite{zhang2020long} \textcolor{black}{for asymptotic keys, while under finite-size effects $\beta$ requirements are higher of 97.4\%~\cite{mani2021multiedge}.}  

For a 100 km fiber link with a total loss of 17 dB, the system achieves similar performance in the asymptotic regime for both cases (blue and red points), owing to comparable excess noise levels. In the finite-size regime, a positive key rate of \textcolor{black}{$5.2\times 10^{-5}$} bits/channel use is achieved at the same 17 dB loss with classical channels ON by increasing the block size to $1.6\times10^9$ \textcolor{black}{(and $\beta$ to 97.4\%)},  considering Gaussian confidence intervals and a failure probability of $10^{-10}$ \cite{ruppert2014long, hajomer2024long}. 

By optimizing $V_M$ with respect to phase noise \cite{hajomer2024long} and a range of attainable reconciliation efficiencies $\beta$, the transmission distance extends to 120 km (20.17 dB loss), achieving a positive key rate in the asymptotic regime (orange point). Given the observed excess noise performance, theoretical simulations (orange solid line) suggest that the total loss budget can extend beyond 21 dB. The tolerable loss can also be translated into large number of network users connected via trusted broadcasting protocol \cite{hajomer2024continuous}. Given such level of noise, up to four users can simultaneously establish a secure key with transmitter distanced by 100 km of ultra-low-loss optical fiber, or more than twenty users at half the distance from transmitter. 

For actual key generation over a 120 km link, we perform error correction with $\beta$ of 93.73\% and a FER of 30\%. At a symbol rate of 125 MBaud, the achieved key rate was 48.3 kbit/s in the asymptotic regime. However, a finite-size key can also be obtained by increasing the block size and improving the efficiency of information reconciliation. Fig.~\ref{fig:100 km}(c) illustrates the minimum required block size $N$ ($\times 10^9$ symbols) and reconciliation efficiency necessary to achieve a finite-size key under the given system parameters (for $D=120$ km in Table.~\ref{tab:1}). At least $5\times 10^9$ symbols are required to reach a positive key with almost perfect efficiency. However, with attainable values $\beta \leq 97\%$ \cite{zhang2020long, mani2021multiedge, hajomer2022modulation} more than \textcolor{black}{12} blocks ($10^9$ each) are needed to reach a positive key rate.


To benchmark our system, we compare its performance with a commercial decoy-state BB84 DV-QKD system operating at 1550 nm, which employs a free-running single-photon detector over a 100 km fiber link. The system generated a secret key rate of 0.88 kbit/s  in the absence of classical channels. However, when the CWDM system was enabled, the key rate dropped to zero. Even after removing the most significant interfering channels (1510 nm, 1530 nm and 1570 nm), the key rate remained zero. 

While DV-QKD systems have demonstrated functionality in noisy CWDM environments through temporal and strong spectral filtering~\cite{townsend1997simultaneous, eraerds2010quantum, patel2012coexistence, wang2017long, gavignet2023co}, our results highlight that CV-QKD provides a plug-and-play solution for long-haul optical links without requiring additional filtering techniques or specific wavelength allocation.

In this work, we demonstrated the longest-distance LLO CV-QKD system coexisting with a fully populated CWDM system over 120 km (20.17 dB loss) with a typical launched power of 1 mW. Importantly, the presence of classical communication has absolutely no impact on the key rate. This achievement is made possible by the inherent properties of the LLO, which serves as an ideal mode filter, and by optimizing the modulation variance to mitigate excess noise due to phase noise~\cite{hajomer2024long}.

One area for improvement in the current implementation is the consideration of composable security, which is currently limited by the block size and the efficiency of information reconciliation~\cite{hajomer2024long}. To enable larger block sizes, the symbol rate can be increased by employing faster detectors, DACs, and ADCs. Moreover, the development of efficient MET-LDPC codes can significantly enhance error correction performance. Using CV-QKD over deployed fiber channels~\cite{williams2024field} also requires addressing polarization drift and clock synchronization, both of which can be effectively managed using digital methods that offer greater flexibility, scalability, and hardware simplicity compared to analog or optical alternatives~\cite{usenko2025continuous}.

 Our results not only demonstrate the longest fiber LLO CV-QKD system to date but, more importantly, establish that CV-QKD can serve as a plug-and-play solution for long-haul optical networks, paving the way for the large-scale deployment of quantum-safe communication.

\begin{backmatter}

\bmsection{Acknowledgments} We thank OFS optics for providing the SCUBA125 fiber for this experiment, Mircea Balauroiu for assistance in operating the DV-QKD system, Naja L.Nysom for conducting error correction, Lucas N. Faria for operating the SNSPDs, and Akash nag Oruganti for discussions on finite-size security. AAEH, ULA and TG acknowledge support from the Danish National Research Foundation, Center for Macroscopic Quantum States (bigQ, DNRF142). This project was funded within the QuantERA II Programme (project CVSTAR) that has received funding from the European Union’s Horizon 2020 research and innovation programme under Grant Agreement No 101017733. ID acknowledges support from the project 22-28254O of the Czech Science Foundation. VCU acknowledges project 21-44815L of the Czech Science Foundation and project CZ.02.01.01/00/22\_008/0004649 (QUEENTEC) of the Czech MEYS. We acknowledge support from  European Union’s Horizon Europe research and innovation programme under the project ``Quantum Security Networks Partnership'' (QSNP, grant agreement no. 101114043), from the European Union’s Digital Europe programme (QCI.DK, grant agreement no. 101091659), and from Innovation Fund Denmark (CyberQ, grant agreement no. 3200-00035B).

\bmsection{Data availability} Data underlying the results presented in this paper are available from the authors upon reasonable request.

\smallskip

\bmsection{Disclosures} The authors declare no conflicts of interest.

\bigskip

\end{backmatter}

\bibliography{sample}

\bibliographyfullrefs{sample}


\ifthenelse{\equal{\journalref}{aop}}{%
\section*{Author Biographies}
\begingroup
\setlength\intextsep{0pt}
\begin{minipage}[t][6.3cm][t]{1.0\textwidth} 
  \begin{wrapfigure}{L}{0.25\textwidth}
    \includegraphics[width=0.25\textwidth]{john_smith.eps}
  \end{wrapfigure}
  \noindent
  {\bfseries John Smith} received his BSc (Mathematics) in 2000 from The University of Maryland. His research interests include lasers and optics.
\end{minipage}
\begin{minipage}{1.0\textwidth}
  \begin{wrapfigure}{L}{0.25\textwidth}
    \includegraphics[width=0.25\textwidth]{alice_smith.eps}
  \end{wrapfigure}
  \noindent
  {\bfseries Alice Smith} also received her BSc (Mathematics) in 2000 from The University of Maryland. Her research interests also include lasers and optics.
\end{minipage}
\endgroup
}{}

\end{document}